\documentclass[aps,prl,reprint,superscriptaddress,longbibliography]{revtex4-1}
\usepackage{graphicx}
\usepackage{amsmath}
\usepackage{mathcomp}
\usepackage{textcomp}

\begin{document}


\title{Mechanical Decoupling of Quantum Emitters in Hexagonal Boron Nitride from Low-Energy Phonon Modes}


\author{Michael Hoese}
\affiliation{Institute for Quantum Optics, Ulm University, D-89081 Ulm, Germany}

\author{Prithvi Reddy}
\affiliation{Laser Physics Centre, Research School of Physics and Engineering, Australian National University, Canberra, ACT 0200, Australia}

\author{Andreas Dietrich}
\affiliation{Institute for Quantum Optics, Ulm University, D-89081 Ulm, Germany}
%

\author{Michael K. Koch}
\affiliation{Institute for Quantum Optics, Ulm University, D-89081 Ulm, Germany}

\author{Konstantin G. Fehler}
\affiliation{Institute for Quantum Optics, Ulm University, D-89081 Ulm, Germany}
\affiliation{Center for Integrated Quantum Science and Technology (IQst), Ulm University, D-89081 Ulm, Germany}

\author{Marcus W. Doherty}
\affiliation{Laser Physics Centre, Research School of Physics and Engineering, Australian National University, Canberra, ACT 0200, Australia}

\author{Alexander Kubanek}
\email[Corresponding author: ]{alexander.kubanek@uni-ulm.de}
\affiliation{Institute for Quantum Optics, Ulm University, D-89081 Ulm, Germany}
\affiliation{Center for Integrated Quantum Science and Technology (IQst), Ulm University, D-89081 Ulm, Germany}


\date{\today}

\begin{abstract}
Quantum emitters in hexagonal Boron Nitride (hBN) were recently reported to hold a homogeneous linewidth according to the Fourier-Transform limit up to room temperature. This unusual observation was traced back to decoupling from in-plane phonon modes which can arise if the emitter is located between two planes of the hBN host material. In this work, we investigate the origins for the mechanical decoupling. Improved sample preparation enabled a reduced background and a 70-fold decrease of spectral diffusion which was so far the major drawback of defect center in hBN and allowed us to reveal a gap in the electron-phonon spectral density for low phonon frequencies. This decoupling from phonons persists at room temperature and explains the observed Fourier Transform limited lines up to 300K. Furthermore, we investigate the dipole emission directionality and show a preferred photon emission through the side of the hBN flakes supporting the claim for an out-of-plane distortion of the defect center. Our work lays the foundation to a deeper understanding of the underlying physics for the persistence of Fourier-Transform limit lines up to room temperature. It furthermore provides a description on how to identify the mechanically isolated emitter within the large number of defect centers in hBN. Therefore, it paves the way for quantum optics applications with defect centers in hBN at room temperature.
\end{abstract}


\maketitle


\section{Introduction}

Fourier-Transform limited (FTL) transitions in atomic systems are among the most crucial ingredients for many quantum optics experiments \cite{Legero2004, Moehring2007, Lettow2010, Togan2010, Sipahigil2012, Ritter2012, Bernien2013, Sipahigil2014}. Furthermore, they might become key building blocks for future large-scale quantum networks \cite{Kimble2008, Sipahigil2016, Wehner2018}. Until recently, the systems exhibiting such transitions, where the spectral width is solely determined by the excited state lifetime, can be divided in two categories. First, atomic systems such as neutral atoms \cite{Kuhn2002, Darquie2005, Thompson2006} or ions \cite{Maunz2007, Barros2009, Almendros2009} that operate at room temperature but with a large technical overhead and with limited rates. Second, atom-like solid state systems that can be operated in compact setups and at high rates but with the requirement for cryogenic temperatures in order to suppress electron-phonon interaction with the solid-state environment \cite{Sipahigil2012, Sipahigil2014, Sapienza2015, Kuhlmann2015, Jahn2015, Chu2017, Bhaskar2017, Dietrich2018}.

In the past few years, a new atom-like system has emerged, namely defect centers in hexagonal Boron Nitrite (hBN) \cite{Tran2016, Toth2019}. The emitter evidence extraordinary optical \cite{Grosso2017, Noh2018, Proscia2018, Tran2018, Dietrich2018, Konthasinghe2019, Nikolay2019} and spin \cite{Exarhos2019, Gottscholl2019, Chejanovsky2019} properties. In addition, first steps towards integrating quantum emitters in hBN into photonic devices were successfully tested \cite{Schell2017, Tran2017, Kim2018, Vogl2019}. A very unique and remarkable feature is the persistence of FTL lines up to room temperature \cite{Dietrich2019}. The unusual persistence of the optical transition against interaction with its solid-state environment was traced back to decoupling from in-plane phonon modes under resonant excitation. However, besides the introduction of a consistent model the experimental proof for the mechanical decoupling is lacking.

In this work, we shed light on new physics based on a quantum emitter which is trapped between two layers of the hBN host. We explain the underlying mechanism of the mechanical decoupling by a systematic spectroscopic study of the excitation and emission properties. At low temperatures we extract the electron-phonon spectral density at low frequencies and examine the persistence of the mechanical isolation up to room temperature. We embed our observations into a consistent theoretical model. In the last paragraph, we analyze the topology of different hBN flakes via AFM microscopy and confirm that all defect centers which are mechanical decoupled emit photons in parallel to the hBN layer orientation. The emission directionality is consistent with a dipole that is distorted out-of-plane. Our work not only gives valuable insights into the new physics of mechanical isolation of defect center in hBN. It also gives a guideline to the community on how to identify those emitters. We therefore contribute important steps to utilize defect center in hBN in quantum optics applications that can be operated at room temperature.

\section{Results}
\subsection{Spectral properties}

\begin{figure*}[]
\includegraphics[scale=1.0]{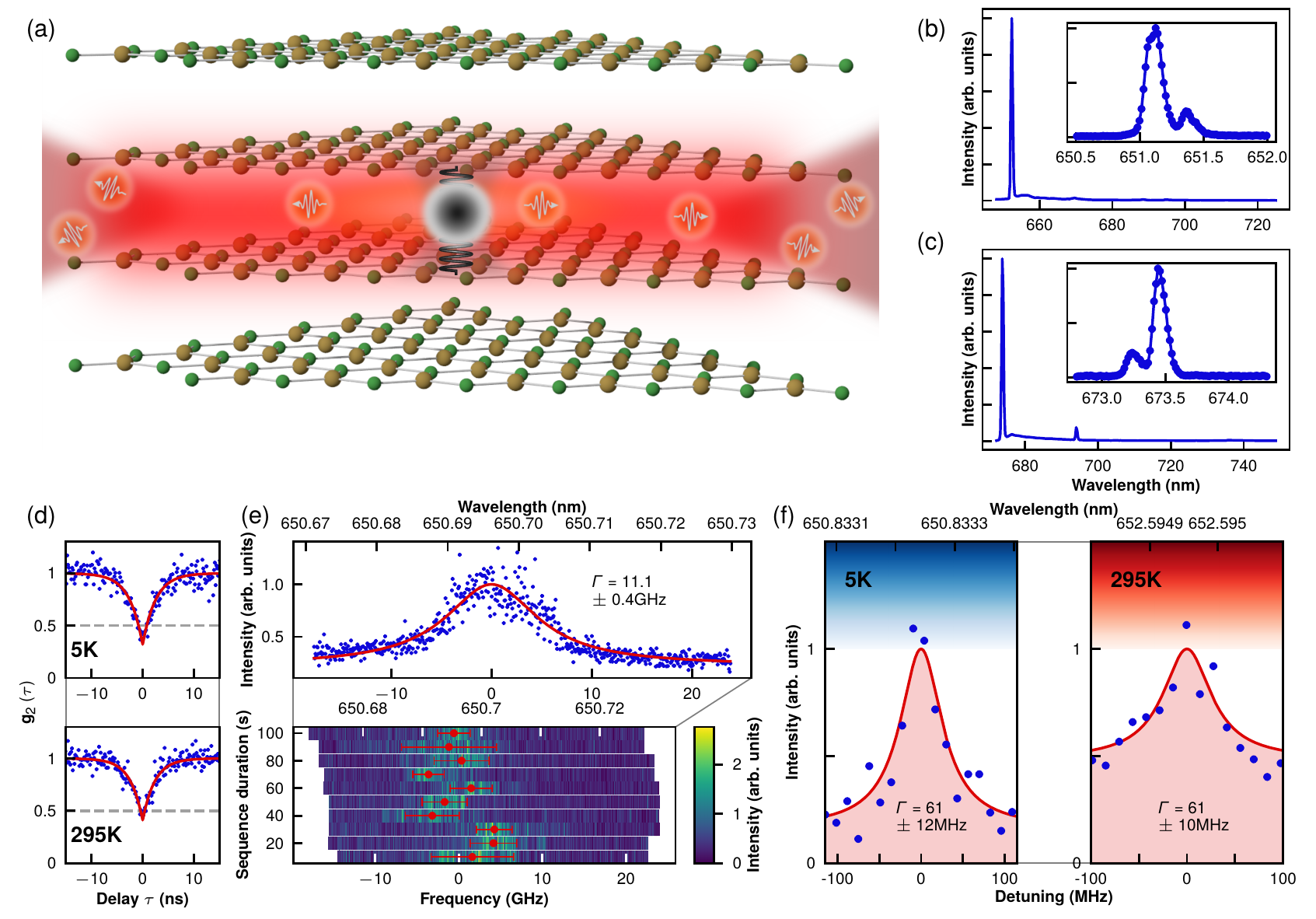}
\caption{\label{Fig:Model-Spectra-Overview}(a) Illustrative drawing of quantum emitters in hBN. (b) PL spectrum of emitter A. Emitter A yields an intense ZPL at 651nm with low background. The inset shows the ZPL in high resolution, thereby revealing two distinct peaks that we identify as independent ZPLs. Since one peak is clearly dominating, the latter one can be treated as background. (c) PL spectrum of emitter B. The PL spectrum of emitter B is dominated by a ZPL at 673nm. The ZPL in high resolution (see inset) shows a dominant ZPL together with a second, less intense peak compared to emitter A. (d) Correlation function of emitter A. The correlation function of emitter A with ZPL at 651nm at 5K and 295K is plotted here (blue dots), together with a fit to the data (red curve), revealing single quantum emitter characteristics with a dip at 0ns delay below 0.5 at both temperatures. The grey lines mark the threshold at 0.5 for single photon emitters. (e) Inhomogeneous linewidth of emitter A. Here, we illustrate 10 subsequent scans over the ZPL resonance. The total inhomogeneous linewidth with a Lorentzian fit to the data is shown in the top panel. Below, we plot the single scans in color scale. The red dots and error bars denote the line positions and linewidths resulting from Lorentzian fits, respectively. (f) Lifetime limited homogeneous lines at 5K and room temperature. Scanning the excitation laser wavelength over the ZPL resonance reveals lifetime limited linewidths when fitting a Lorentzian (red curve) to the measurement data (blue dots) at both, cryogenic temperatures (top panel) and room temperature (bottom panel).}
\end{figure*}

As illustrated in figure \ref{Fig:Model-Spectra-Overview}(a), our proposed model for explaining the persistence of Fourier-Transform limited lines up to room temperature rests upon mechanical decoupling of the optical transition from phonon modes in the hBN lattice \cite{Dietrich2019}. Emitters located between hBN layers may be the cause of this decoupling because their orbitals are less susceptible to in-plane phonon distortions. In order to understand the underlying mechanism, we first measure the photoluminescence (PL) spectrum as input for a phonon sideband decomposition model. This allowed us to extract the electron-phonon spectral density and associate its resonances to hBN phonon modes.  We then in turn utilize resonant excitation of phonon resonances to investigate the coupling of the modes under resonant drive. Since our laser system (Sirah Matisse 2 DS) operates in a wavelength range limited to 618 - 671 nm, we focus on two emitters where one, emitter A (see fig. \ref{Fig:Model-Spectra-Overview}(b)), can be excited resonantly and near resonantly to the zero phonon line (ZPL) and the second one, emitter B (see fig. \ref{Fig:Model-Spectra-Overview}(c)), can be excited further away (up to 56 nm) from the ZPL via the optical phonon mode. Emitter A features a ZPL at a wavelength of 651nm. Its PL spectrum shows a low background fluorescence and high signal intensity. We measure the second order autocorrelation function in Hanbury-Brown and Twiss (HBT) configuration to verify the single photon character of our emitters. The thereby arising relatively high value of $g_2\left(0\right)=0.32 \pm 0.03$ (see fig. \ref{Fig:Model-Spectra-Overview}(d)) originates from a second, uncorrelated ZPL, as reported previously in \cite{Bommer2019}. At room temperature, this value rises to $g_2\left(0\right)=0.41 \pm 0.02$ due to higher background fluorescence. As observed in the PL spectrum at 5K, the latter ZPL is only 0.25 nm detuned and five times less intense than the main ZPL (see fig. \ref{Fig:Model-Spectra-Overview}(b)). Nevertheless, the single photon character is clearly proven for 5K and 295K since $g_2\left(0\right)<0.5$. Fitting the second order correlation function
\begin{equation}
g_2\left(\tau\right) = 1-a\exp\left(\frac{|\tau|}{\tau_0}\right)
\end{equation}
to data from a correlation measurement yields a natural linewidth of $62.9 \pm 3.1\mathrm{MHz}$ at cryogenic and $66.8 \pm 4.2\mathrm{MHz}$ at room temperature, in line with previous reports \cite{Tran2016a, Sontheimer2017, Tran2018}. The excitation power is kept well below saturation in order to avoid power broadening. For further analysis it is important to note that the contribution of the second emitter to the phonon sideband (PSB) emission is small. Next, we perform resonant photo luminescence excitation spectroscopy (PLE) by scanning the laser frequency over the emitter transition frequency with a tunable dye laser. The Lorentzian fit to the resonance, as illustrated in \ref{Fig:Model-Spectra-Overview}(e), yields an inhomogeneous linewidth at 5K of $11 \pm 0.4\mathrm{GHz}$ hinting to residual spectral diffusion. Compared to earlier works \cite{Dietrich2019}, we iteratively improved our sample preparation process, resulting in clean emitter spectra and low background fluorescence. Furthermore, the cleaner substrate with less surface disturbances reduces spectral broadening and stabilizes the emitter. Thus, we could improve the inhomogeneous linewidth by a factor of 70 at comparable excitation powers of few $\mu\mathrm{W}$. For individual scans we encounter a homogeneous linewidth within the FTL of $62.9 \pm 3.1\mathrm{MHz}$. An exemplary scan at cryogenic temperature (5K), reveals a homogeneous linewidth of $61 \pm 12\mathrm{MHz}$, as shown in Fig. \ref{Fig:Model-Spectra-Overview}(f). At room temperature (293K), a scan with a measured linewidth of $61 \pm 10\mathrm{MHz}$ (see Fig. \ref{Fig:Model-Spectra-Overview}(f)) confirms the FTL of the homogeneous linewidth.

\begin{figure*}[]
\includegraphics[scale=1.0]{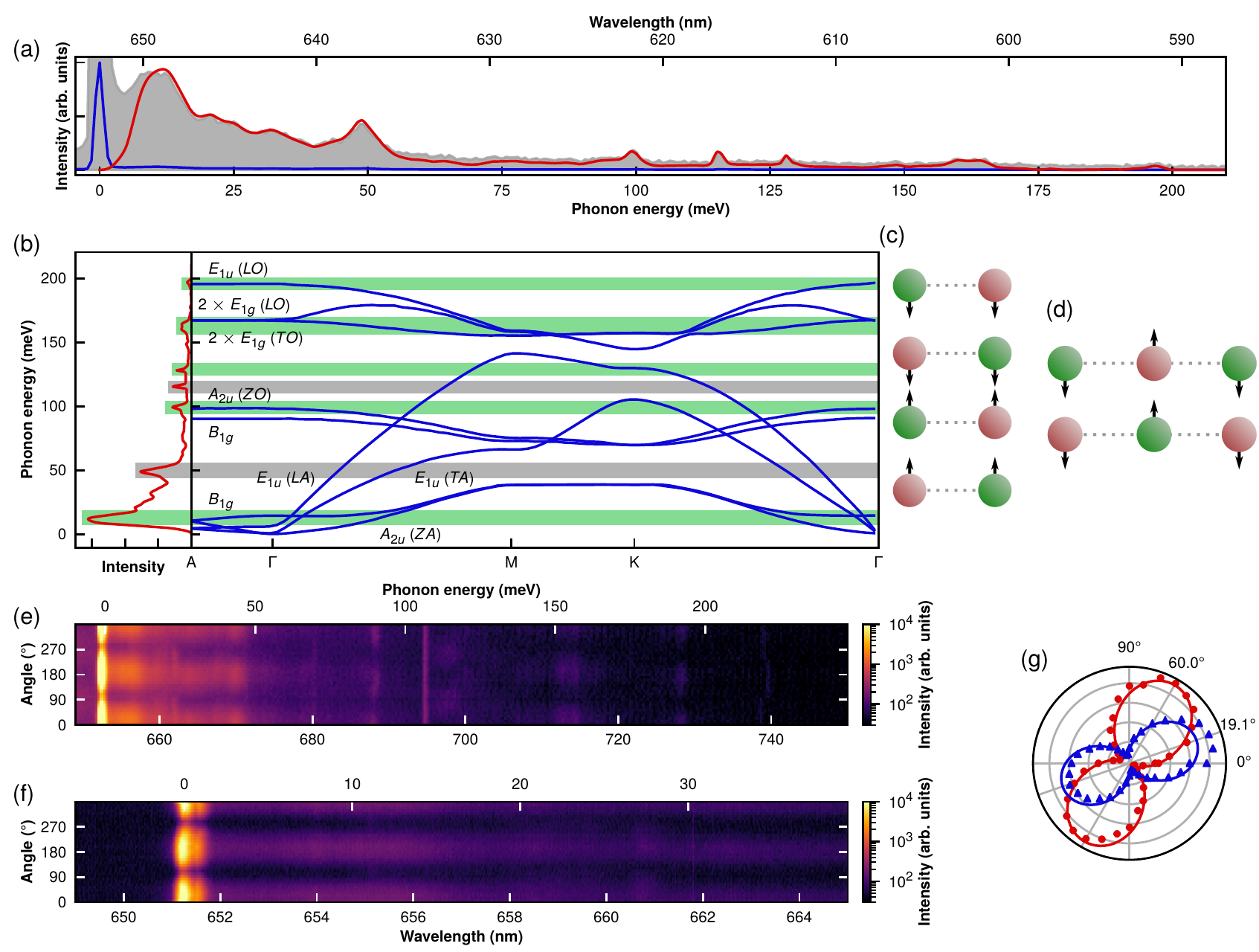}
\caption{\label{Fig:Polarization-SA01-Emitter66} (a) PL spectrum. Here, we display the full spectrum (red curve) of emitter A. For better visibility, the grey background illustrates the magnified PSB. For comparison, we plot the one-phonon band, derived from PSB decomposition in red. (b) Comparison of the one-phonon band and the phonon band structure. Each band in the band structure is labelled by its symmetry at the $\Gamma$-point. The green highlighted peaks in the one-phonon band (plotted in red at the right) are coincident with bands leveling off at high symmetry points in the band structure. The grey bars indicate peaks that are not coincident with any features in the band structure. (c) The $B_{1g}$/$A_{2u}$ modes at the $A$-point. Here we illustrate the atom motion resulting in the $B_{1g}$/$A_{2u}$ bands at the $A$-point. (d) The $A_{2u}$ mode at the $\Gamma$-point. This out-of-plane mode is sketched here with arrows denoting atom motion. (e) Polarization dependent PL. The image shows a polarization dependent PL spectroscopy with the logarithmic color scale at the ZPL at 651nm being cut off for better total visibility. (f) Polarization dependent PL in high resolution. Here, we plot polarization dependent PL of the PSB close to the ZPL with high resolution spectra. For better visibility, the colorscale is logarithmic. (g) 532nm absorption and emission polarization. The red data points together with the red fit denote the polarization of the 532nm laser light absorbed for PL spectroscopy. The polarization of light emitted from the emitter is plotted with blue dots and a blue curve marking the measurement data and the fit, respectively. Both measurements reveal a clearly distinct orientation of the 532nm absorption and the emission dipole.}
\end{figure*}


\subsection{Phonon sideband decomposition}

Now, we study the PL spectrum of emitter A in detail to identify the defect's electron-phonon coupling mechanisms. In figure \ref{Fig:Polarization-SA01-Emitter66}(a), the blue curve shows the defect's bright ZPL emission. The grey shaded intensity shows a magnified view of the PSB which highlights the phonon features. We applied a PSB decomposition method which adopts a linear symmetric mode model \cite{Davies1974, Kehayias2013, Balasubramanian2019} to describe the PSB in terms defect's electron-phonon spectral density, also referred to as the one-phonon band. The one-phonon band represents all defect processes involving the creation and/or annihilation of a single phonon. Using our method we accurately decomposed the PSB into its one-phonon band. Further details of the decomposition process are included in the supplemental material \footnote{see supplemental material}.  Since the band can be accurately modeled in terms of its one-phonon band, we can assume the absence of strong vibronic interactions that involve non-symmetric modes or introduce anharmonic effects, like Jahn-Teller.

Next, we compare the intensities of the one-phonon band features to the intrinsic phonon modes of hBN to identify the defect's electron-phonon processes. Here we are assuming that the emitter does not significantly perturb the modes of pristine hBN. Features in the one-phonon band correspond to frequencies of high mode density and/or where there is strong coupling to the defect orbitals. The one-phonon band is plotted against the phonon band structure of pristine hBN as shown in figure \ref{Fig:Polarization-SA01-Emitter66}(b). The modes that couple to the defect are highlighted in green. The one-phonon band's largest feature is a broad peak centered at 11meV. The prominence of this feature indicates that, at this frequency, two things are occurring: a high density of modes and strong coupling to the defect. The feature is coincident with the leveling out of the $B_{1g}$ and $A_{2u}$ bands at the $A$-point. These are acoustic modes in the out-of-plane direction, their qualitative atomic displacements are shown in Fig. \ref{Fig:Polarization-SA01-Emitter66}(c). The $A$-point corresponds to out-of-plane wave-vector at the edge of the Brillouin zone. This means that $A$-point phonons also result in maximum displacement between equivalent atoms in neighboring unit cells in the inter-plane direction. There is a sharp feature at 50meV that does not coincide with any critical points in the band-structure, indicating that it is a local mode. The next green peak corresponds to the $A_{2u}$ mode at the $\Gamma$-point, which is the out-of-plane optical mode in Fig. \ref{Fig:Polarization-SA01-Emitter66}(d).
The feature at 115meV does not coincide with any band structure detail, since it stems from the Raman line of chromium defects in the sapphire substrate and not from the emitter itself.

The remaining features of the one-phonon band imply that the defect weakly couples to the following: (1) the longitudinal acoustic $E_{1u}$ mode at the $K$-point, (2) potentially multiple transverse modes, and (3) the longitudinal optical $E_{1u}$ mode at the $\Gamma$-point. These modes correspond to in-plane displacements between neighboring unit cells of hBN. The microscopic displacements that can be associated to the last three features are quite complicated. As a result, it is not immediately obvious how they may interact with the defect. Nevertheless, these observations may still provide some insight as we develop our understanding of the defect.

Given the defect’s strong coupling to modes that result in out-of-plane displacements, we conclude that the orbitals of the defect exist between the layers of hBN. This is consistent with the picture presented in \cite{Dietrich2019} that an out-of-plane distortion has led to decoupling from in-plane phonons. We generalize the result to other interplane defects like a trapped atom near an impurity or vacancy.

The polarization-dependent PL spectrum shown in figures \ref{Fig:Polarization-SA01-Emitter66}(e) and (f), exhibits the polarization of the full PSB. Light emitted from the ZPL shows a distinct polarization contrast. Thereto, the polarization contrast for the off-resonant excitation at 532nm is rotated by 41\textdegree (see Fig \ref{Fig:Polarization-SA01-Emitter66}(g)), as observed previously \cite{Jungwirth2016} and hinting towards an additional excited state. The polarization of both the optical phonon mode at 710nm and the low-energy acoustic phonon modes are aligned to the ZPL emission. We conclude that the involvement of phonons in the emission process does not alter the polarization contrast. This is consistent with the defect coupling to linear symmetric modes.


\begin{figure*}[]
\includegraphics[scale=1.0]{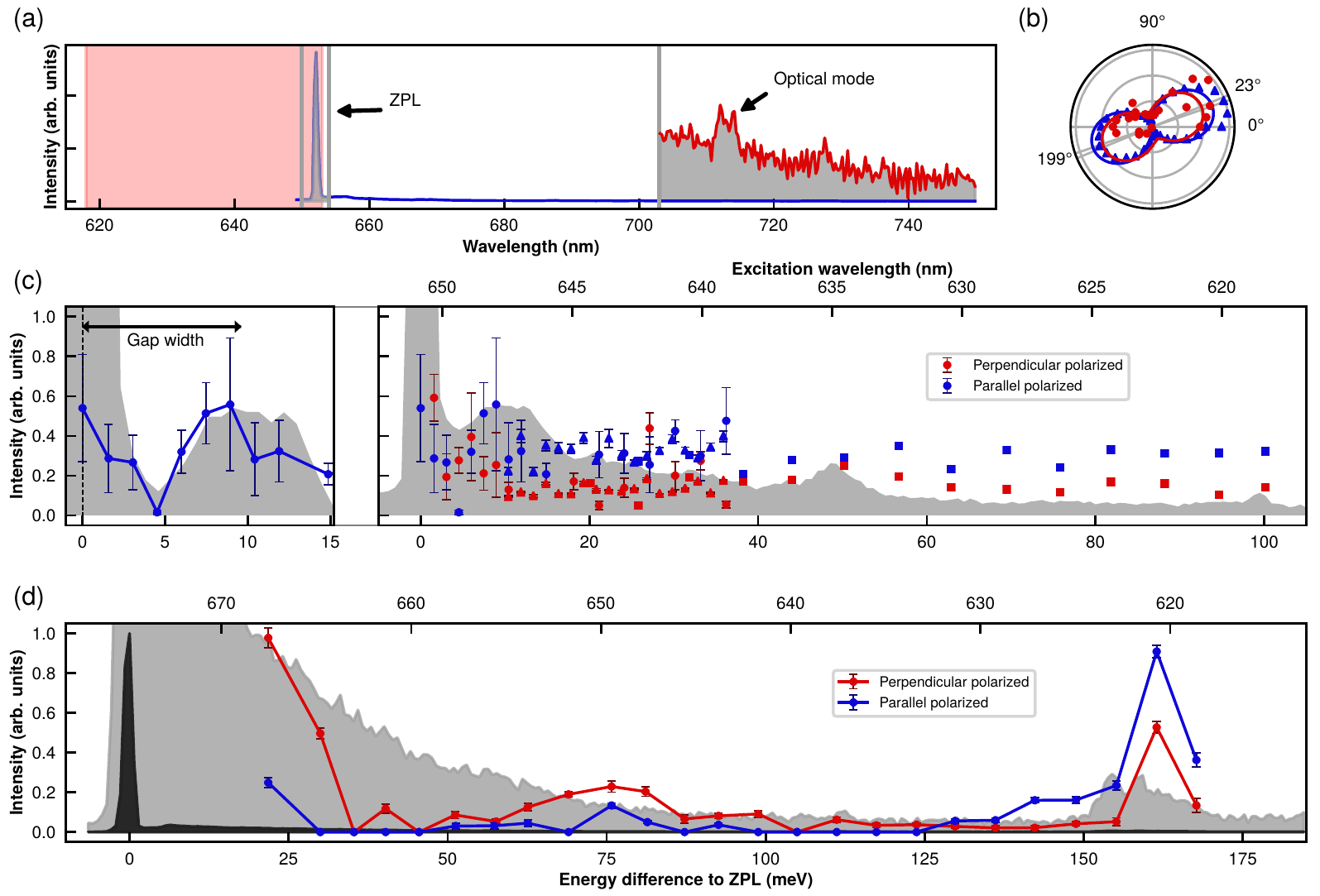}
\caption{\label{Fig:Em66-ExcitationPSB}(a) PLE measurement scheme for excitation via PSB. The blue curve shows a PL spectrum of emitter A with a ZPL at 651nm. The magnified PSB is displayed in red. The excitation wavelengths that we use for PLE are shaded in red. For detection, we acquire spectra and use either the integral over the ZPL or the integral over the PSB as measure for the intensity. Both integration areas are shaded in grey. (b) Polarization of 1st phonon mode. We excite emitter A with 6.8meV detuning from the ZPL. We fit the measurement data (red points) with a sine (red curve), yielding a polarization angle of 23\textdegree. For comparison, we plot the emission polarization of the ZPL in blue. (c) Polarization dependent PLE at the excitation PSB of emitter A. In the right panel, blue and red dots represent results for excitation with parallel and perpendicular polarized light, respectively. Circles and triangles denote two different measurement runs, using the optical mode for detection. Hereby, we acquire spectra with long exposure time and integrate the spectrum over the PSB. Squares extend the measurement to higher energy differences with data extracted from the ZPL integral. The grey shaded spectrum in the background is a mirrored PL spectrum for comparison with PLE measurements at the excitation PSB. In the left panel, we illustrate the extracted gap width. We determine the peak maxima of the ZPL and of the first phonon mode and equalize the difference with the gap width. For better illustration we rescale the intensity of the PL measurement demonstrating the consistency of the gap size in the PL and PLE measurement. (d) PLE at the excitation PSB up to the optical phonon mode. Here, we use emitter B with a  ZPL at 673nm. The blue points denote the results for parallel polarization and red for perpendicular, respectively. All measurement values are extracted by integration over the ZPL. The line connecting the measurement values serves as guide to the eye. The dark grey shaded spectrum in the background is the PL spectrum of the emitter mirrored at the ZPL. For better visibility, we plotted the magnified PSB in PL in lighter grey.}
\end{figure*}

\subsection{Coupling to excited state phonon sideband}

Next, we probe the electron-phonon coupling strength of individual phonon resonances. For this purpose we utilize wavelength-dependent PLE spectroscopy. We operate the experiment in two different modes. First, we integrate the spectrum over the optical phonon sideband from 705nm to 800nm, as illustrated in Fig. \ref{Fig:Em66-ExcitationPSB}(a), in order to detect the fluorescence when exciting close to the ZPL. Second, we use the integral of a Gaussian fit to the ZPL as intensity measure when exciting further away from the ZPL. Thereby, we can measure the excitation strength over a large frequency range and compare it to the PL signal when adjusting the scaling of the emitter intensity. The first observation is that we can clearly excite the first phonon feature, which can be associated to the first ZA mode, 8 meV detuned from the ZPL. The polarization contrast of the excitation on-resonance with the acoustic mode yields the polarization-dependence as depicted in Fig. \ref{Fig:Em66-ExcitationPSB}(b). In contrast to the polarization of the off-resonant excitation at 532nm we now observe the excitation polarization aligned to the ZPL emission polarization. We conclude that we are now addressing the same excited state with a ZA-phonon-assisted excitation that preserves the polarization. Figure \ref{Fig:Em66-ExcitationPSB}(c) unfolds the polarization- and wavelength-dependent electron-phonon coupling across the complete phonon spectrum. We note, that there is also a weak excitation probability on-resonance with the ZA mode but with perpendicular polarization with respect to the ZPL emission. We interpret this observation as residual excitation of the second, much weaker ZPL. Since that second ZPL of both defects is detuned by around 0.8meV (see insets in figs. \ref{Fig:Model-Spectra-Overview}(b) and (c)), we do not resolve the splitting here. Now, we want to investigate the low-frequency gap between the ZPL and the acoustic phonon branches and therefore focus on the data polarized parallel to the ZPL emission. The fluorescence intensity and therefore the electron-phonon coupling strength displays a clear minimum at about 5meV detuned from the ZPL confirming our findings in PL spectroscopy. The maximum on-resonance with the first acoustic phonon energy at around 8 meV detuned from the ZPL resonance at 0 meV also agrees with the results from the PL spectrum. The left panel in figure \ref{Fig:Em66-ExcitationPSB}(c) illustrates the gap width for both, PL and PLE data. Therefore, we determine the peak positions of ZPL and PSB, respectively, and define the gap size as their difference. The figure illustrates the consistency of the gap width between PL and PLE measurement.


We now finalize our study of the electron-phonon coupling strength by investigating the optical phonon mode. Due to electromagnetic coupling between the optical transition and the optical mode we expect electron-phonon coupling even for an emitter that is displaced out-of-plane of the hBN layer. Due to the limited tuning range of our laser system we use a second emitter B with a ZPL at 673nm to probe excitation via the optical phonon mode. The PL spectrum and the wavelength-dependent excitation is shown in figure \ref{Fig:Em66-ExcitationPSB}(d). The emitter can be excited efficiently via the optical phonon mode showing a distinct resonance at 165meV detuned from the ZPL and in agreement with previous results \cite{Wigger2019}. The polarization of the excitation laser is again aligned with the polarization of the ZPL emission confirming that the phonon-assisted excitation does not alter the  polarization properties. Please note, that also in this case we do observe a second ZPL with an intensity of approximately half of the first ZPL. That ZPL could explain the residual excitation probability perpendicular to the ZPL emission. Furthermore, please note that we observe signs of additional modes in excitation at a detuning of approximately 75 meV from the ZPL. However, assigning these resonances to modes in the PL spectrum is not distinct.

\begin{figure}[]
\includegraphics[scale=1.0]{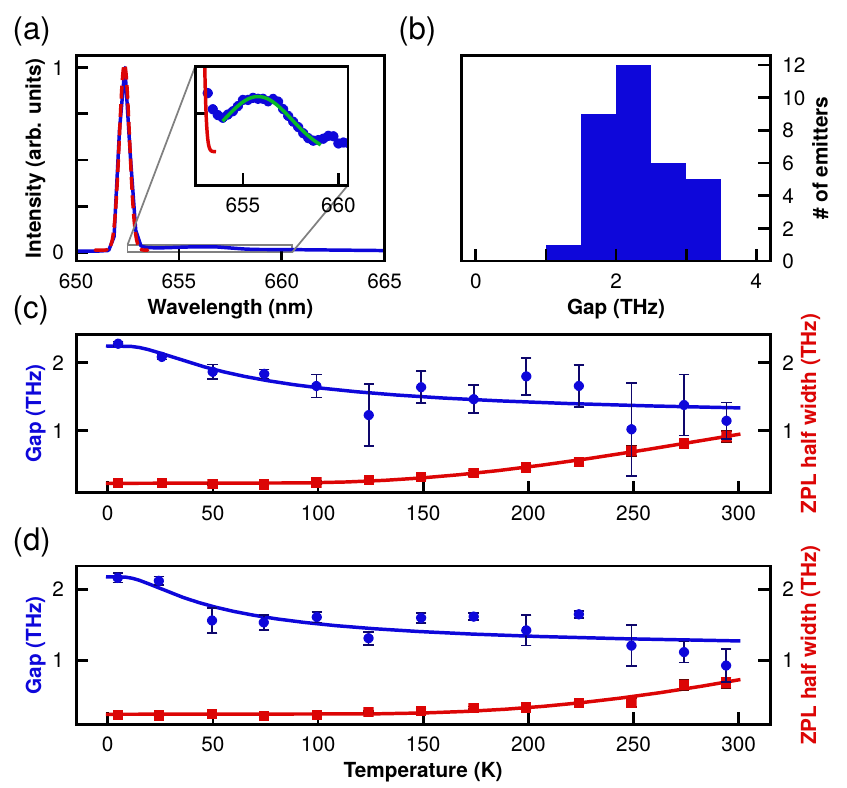}
\caption{\label{Fig:GapSizeTdependent}(a) Gap size analysis procedure. For evaluation of the gap size, we fit a Gaussian (dashed red curve) to the ZPL in the PL spectrum (blue curve) and to the PSB. The inset shows the fitted PSB (green curve) with the PL spectrum (blue dots). (b) Gap size distribution. We determine the gap size of multiple emitters at 5K and display the abundance distribution, peaking at 2THZ. (c) Evolution of the gap size. Here, we measured the gap size (blue points) and the half width of the ZPL (red squares) for different temperatures. The data here stems from emitter A, which we used for resonant PLE and PLE at the excitation PSB. For extrapolation, we assumed proportionality of both measurement values to the Boltzmann factor. The fitted curves are plotted in blue and red for the according parameters, respectively. (d) Evolution of the gap size. As in fig. (c), we depict the gap data for emitter B using the same labeling.}
\end{figure}

\subsection{Temperature dependent gap size}

In the following, we focus on the temperature dependence of the gap in the electron-phonon spectral density and its persistence up to room temperature. We extrapolate the gap size by fitting a Gaussian to the ZPL and to the first acoustic phonon feature and determine the gap size as distance between both peaks. An example is illustrated in Fig. \ref{Fig:GapSizeTdependent}(a). A characterization of 33 different emitters yields an average gap size of about 2 THz within a distribution of approximately 1 THz, as illustrated in Fig.\ref{Fig:GapSizeTdependent}(b). The histogram is asymmetric and drops faster towards smaller gap size since smaller gap widths are more difficult to resolve. In order to be able to observe the gap its size needs to be larger than the ZPL linewidth. We characterize both, the gap size of emitter A and B for different temperatures from 5 K to 300 K, as plotted in Fig. \ref{Fig:GapSizeTdependent}(c) and (d), respectively and plot it against the ZPL linewidth. With rising temperature the gap size narrows whereas the width of the ZPL increases. The gap of both emitters remains open and observable all the way to room temperature. However, the gap size at 300 K decreases to about half the size at 5 K while the ZPL width increases to almost half the gap size at 300 K. We fit the temperature dependence of both, the ZPL linewidth and the gap size, with the Boltzmann function
\begin{equation}
w = A + B \cdot \exp \left(-\frac{C}{k_\mathrm{B}T} \right),
\end{equation}
with $w$ denoting the gap size or the ZPL width and the fit constants $A$, $B$ and $C$ and extrapolate a gap size of $1.14 \pm 0.24$ THz at 294.2 K with a ratio for the Boltzmann energy of $\frac{hf}{k_BT} = 0.186 \pm 0.039$. This dependence could arise from a temperature-dependent distortion that modifies the electron-phonon couplings and phonon frequencies. For example, this could be a further out-of-plane distortion of an already distorted defect or changes to the configuration of an inter-plane defect. For changes in the electron-phonon coupling to shift the energies of PSB features, there must be a difference in the phonon frequencies between the two electronic levels of the optical transition. This is a higher order effect that is not captured by the linear symmetric mode model and warrants future investigation.

\subsection{AFM flake characterization}

\begin{figure*}[]
\includegraphics[scale=1.0]{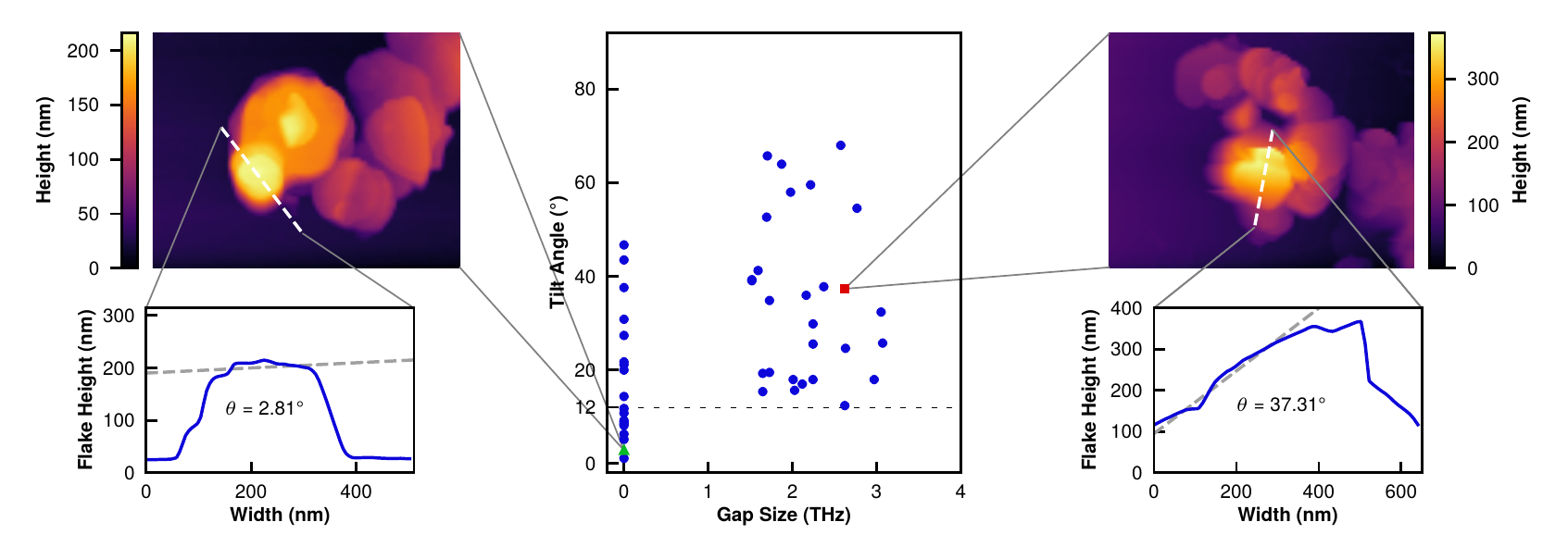}
\caption{\label{Fig:AFMScan} AFM Characterization. Here, we present the key results of an AFM characterization of flakes hosting quantum emitters with and without gap between ZPL and PSB. The central plot depicts the tilt angle of host flakes versus the gap size of the emitters. The dashed black line marks 12° tilt and the grey shaded area highlights the minimum tilt for emitters with gap. On the left side, we illustrate the flake that provides the data of the green triangle in the central graph. The white dashed line in the 3D AFM height profile marks the segment we use for measuring the tilt of the flake. Below, we plot this 2D height profile together with a straight dashed line illustrating the tilt angle determination. In the same manner, we exemplary picture one flake hosting an emitter with gap in the right panel. The 2D height profile reveals the strong tilt of the host flake resulting in the red squared data point.}
\end{figure*}

Finally, we characterize the topology of the hBN flakes with an atomic force microscope (AFM). We compare the flake texture between flakes hosting emitters with and without gap in the electron-phonon density of states. The AFM scan of the height profile reveals the angle of the hBN flake surface with respect to the substrate surface, which we equalize to the hBN layer orientation. Two exemplary scans are shown in figure \ref{Fig:AFMScan}. We now gain insights into the mechanical decoupling mechanism by correlating the tilt angle of the hBN flakes hosting the defect centers with the gap size observed in the PL spectrum. If an out-of-plane distortion of the emitter is the origin for the mechanical decoupling, then all emitters with a gap should only be observable in tilted hBN flakes. In contrast, all emitters without a gap should predominantly be observed in hBN flakes lying flat on the surface corresponding to small tilt angles. This correlation is clearly visible in figure \ref{Fig:AFMScan}. In total 45 emitters are studied. All emitters with no gap in the PL spectrum are observed in hBN flakes with a tilt angle below 50°. In fact every second flake has a tilt of less than 12°. Vice versa, all flakes hosting emitters with a gap in the PL spectrum yield tilt of at least 12°. Possibly, the dipole of emitters with gap is distorted out-of-plane
and the emission tends to spread along the hBN layers. This results in a better 	visibility when the flake is tilted such that the hBN layers and the emitted light points towards the objective for collection. On the other hand, in-plane defects result in gapless emitters. Since their dipoles are also located in-plane the emission is mostly perpendicular to the hBN layers. We observe this class of emitters in flakes with small tilt angles.

\section{Conclusions}

In summary, we confirm that some defect center in hBN exhibit mechanically decoupled electronic transitions under resonant excitation. Improved sample preparation enabled a 70-fold decrease in spectral diffusion and allowed us in this work to observe a gap of around 2 THz in the electron-phonon density of states which remains open up to room temperature. In a systemic study of the coupling to individual modes we developed a model where the emitter has orbitals that exist between the layers of hBN. This model was proposed in reference \cite{Dietrich2019} to explain the persistence of FTL linewidths in resonant excitation up to room temperature. We furthermore categorize common features of all emitters with decoupled optical transitions showing a gap in the PL and wavelength-dependent PLE spectrum between ZPL and first acoustic phonon modes. Our results imply that only multilayer hBN flakes can serve as host for this emitter type. Furthermore, the emission directionality is shifted towards an emission parallel to the hBN layer. Besides new insights into the physics explaining the extraordinary observation of Fourier-transform limited lines at room temperature our work also outlines a catalog that could be used in future experiments to identify mechanically decoupled emitters. Our work therefore opens up new ways to use defect center in hBN for quantum optics applications at room temperature.

\medskip
\begin{acknowledgments}
The project was funded by the Deutsche Forschungsgemeinschaft (DFG, German Research Foundation) - Project number: 398628099. A.K. acknowledges support of the European fund for regional development (EFRE) program Baden-Württemberg. M.W.D. acknowledges support from the Australian Research Council (DE170100169). K.G.F. and A.K. acknowledge support of IQst. M.H. acknowledges support from the Studienstiftung des deutschen Volkes. The AFM was funded by the DFG. We thank Prof. Kay Gottschalk and Frederike Erb for their support. The Qudi software suite \cite{Binder2017} was used for running the experiment setup.
\end{acknowledgments}

\bibliography{ExcPSB}

\end{document}